\documentclass[aps,prx,reprint,superscriptaddress,nofootinbib,longbibliography,floatfix,showkeys]{revtex4-2}

\usepackage[T1]{fontenc}
\usepackage[utf8]{inputenc}
\usepackage{amsmath,amssymb,bm,mathtools}
\usepackage{graphicx}
\usepackage{booktabs}
\usepackage{xcolor}
\usepackage{hyperref}
\usepackage{microtype}
\usepackage{placeins}

\hypersetup{colorlinks=true,linkcolor=blue!55!black,citecolor=blue!55!black,urlcolor=blue!55!black}


\makeatletter
\def\@keys@name{Subject Areas: }
\makeatother

\newcommand{\rr}{\bm r}
\newcommand{\rin}{\rr_{\rm in}}
\newcommand{\rout}{\rr_{\rm out}}

\newcommand{\Mtar}{M_{\rm tar}}
\newcommand{\ctar}{\bm c_{\rm tar}}
\newcommand{\Ldyn}{\mathcal L_{\rm dyn}}
\newcommand{\Lgate}{\mathcal L_{\rm gate}}
\newcommand{\Lreg}{\mathcal L_{\rm reg}}
\newcommand{\LT}{\mathcal L_T}

\newcommand{\Ltotal}{\mathcal L_{\rm total}}
\newcommand{\dd}{\mathrm d}
\newcommand{\appsecref}[1]{Appendix~\hyperref[#1]{\ref*{#1}}}

\newcounter{resultsubpart}[subsection]
\renewcommand{\theresultsubpart}{\arabic{resultsubpart}}

\newcommand{\resultsubpart}[1]{%
  \par\addvspace{0.50\baselineskip}%
  \refstepcounter{resultsubpart}%
  {\centering\itshape \theresultsubpart.\,#1\par}%
  \nobreak\vspace{0.15\baselineskip}%
  \noindent\ignorespaces
}

\DeclareRobustCommand{\panlabel}[2]{\hypertarget{pan.#1.#2}{}}
\newcommand{\panref}[2]{\hyperlink{pan.#1.#2}{(#2)}}

\begin{document}

\title{Evolution-Level Quantum Optimal Control of Single-Qubit Gates with Physics-Informed Neural Networks}

\author{Yao Du}
\affiliation{School of Physics and Information Technology, Shaanxi Normal University, Xi'an 710119, China}
\author{Jian-Jian Cheng}
\affiliation{School of Science, Xi'an University of Posts and Telecommunications, Xi'an 710121, China}
\author{Lin Zhang}
\affiliation{School of Physics and Information Technology, Shaanxi Normal University, Xi'an 710119, China}
\author{Ming-Liang Hu}
\affiliation{School of Science, Xi'an University of Posts and Telecommunications, Xi'an 710121, China}
\author{Xingang Wang}
\email[Corresponding author: ]{wangxg@snnu.edu.cn}
\affiliation{School of Physics and Information Technology, Shaanxi Normal University, Xi'an 710119, China}
\date{\today}

\begin{abstract}
Quantum gate design is often represented as pulse optimization, although the
physical object that implements a gate is the full controlled evolution
generated by the pulse.  Here we use physics-informed neural networks to
represent single-qubit gate design at this evolution level: the control
fields, the Bloch-state trajectories, and the total duration are learned together
under the Bloch equation.  This changes the optimized object from pulse
amplitudes to a differentiable physical process whose structure can be
inspected and refined.  For rotation gates, the optimized evolutions recover
the physical organization expected for bounded single-qubit control, with no
prescribed pulse ansatz or duration scan.  For a geometric gate, the
representation identifies localized bottlenecks in maintaining the geometric
condition and turns this diagnosis into feedback, reducing the residual path
error while preserving high fidelity.  Thus physics-informed learning is used
not only to synthesize gates, but also to make optimized quantum controls
physically readable, diagnosable, and locally refinable.  This process-level
view may be especially useful for adapting gates to hardware-specific,
task-specific, and locally varying experimental constraints.
\end{abstract}


\maketitle

\section{Introduction}

In many complex physical problems, obtaining a high-quality numerical
solution does not automatically mean that the solution has been understood
\cite{Schmidt2009NaturalLaws,Brunton2016SINDy,Iten2020PhysicalConcepts,Cranmer2020SymbolicModels}.
This is especially true in high-dimensional optimization problems, where an
algorithm may efficiently produce an excellent answer while the variables,
constraints, and regularities that organize that answer remain obscure.  In
this sense, solving a problem is not the same as seeing through it.  Quantum
optimal control provides a particularly sharp example of this distinction: a
gate may be realized with high fidelity, yet the controlled process by which
the gate is formed can remain only indirectly understood.

This issue becomes concrete in quantum optimal control, whose goal is to use
externally tunable fields to steer a
quantum system through a desired evolution and implement a target quantum
operation.  For gate-design tasks, high-fidelity control solutions can be
found systematically, and existing methods are highly successful as
numerical solvers, particularly once a time window and a pulse
parametrization have been specified
\cite{Brif2010QuantumControlReview,Glaser2015TrainingCat,Koch2022QuantumTechnologies,Khaneja2005GRAPE,Palao2003KrotovUnitary}.
Yet a high-performing pulse by itself does not necessarily reveal how the
control process is organized.  This distinction becomes important when
control design is expected to move beyond finding a waveform that works in
an ideal simulation.  In experimentally relevant settings, one often needs
controls that remain compatible with hardware limitations, indicate how the
gate is actually formed, and can be systematically refined when calibration
errors, drifts, or local path constraints appear.  Three questions then
arise naturally: How should the gate duration be determined when the
available control strength is finite? What pulse shape physically realizes
the target operation under realistic restrictions? For a geometric gate, how
can one verify that the whole path is genuinely geometric rather than only
correct at the final time? In standard workflows these questions are often
handled separately: the duration is chosen or scanned beforehand, the pulse
is optimized, and trajectory- or path-level properties are checked
afterward.  This separation is useful computationally, but it obscures the
fact that these quantities belong to one controlled evolution.

The physics points to a different organization.  A quantum gate is not a
waveform alone but a controlled dynamical process: the fields generate
trajectories, the trajectories determine the implemented operation, and the
duration sets the time scale on which the dynamics and the target can be
satisfied
\cite{Peirce1988QuantumOCT,Ramakrishna1995Controllability}.  For geometric
gates, the path followed by the state is also part of the operation itself.
These elements are therefore physically coupled.  In the most common
discrete-pulse formulation, however, the complete process is represented
mainly through pulse parameters
\cite{Khaneja2005GRAPE,Caneva2011CRAB,Machnes2018GOAT}: the pulse is the
object directly varied by the optimizer, while the induced trajectories and
path properties are reconstructed or checked afterward.  The decomposition
is therefore not a demand of the physics but a limitation of what a discrete
representation can hold together.  This distinction matters when the
scientific question is not only whether a control field exists, but how the
gate is physically built in time and how that information can be used to
refine the control itself.

This work starts from this point and uses physics-informed neural networks
(PINNs) to construct a continuous representation of controlled gate
dynamics.  PINNs encode a governing equation as a differentiable constraint,
so that the dynamics are represented by differentiable functions and tested
through equation errors along the physical interval
\cite{Raissi2019PINN,Karniadakis2021PIML,Cuomo2022PINNReview}.  The same
idea appears in continuous-depth and universal-differential-equation models,
which treat a dynamical process as a differentiable object rather than a
fixed list of samples \cite{Chen2018NeuralODE,Rackauckas2020UDE}.  In
quantum control, PINN-based studies have already shown that neural
representations can produce high-fidelity state transfers and gates
\cite{Guengordu2022SmoothPINNGates,Norambuena2024PINNQuantumControl,Lauten2025GateDesign,Ullah2024QuantumDissipative}.
The question here is not only whether a high-fidelity gate can be found, but
what changes when the gate-forming evolution itself is made the represented
and optimized object.

Our formulation treats single-qubit gate design as the optimization of a
continuous controlled process.  The optimized object is no longer only a
finite vector of pulse amplitudes.  Instead, one trainable representation
generates the control fields and the relevant trajectories as continuous
functions of physical time, while violations of the Bloch equation, the gate
requirement, the time cost, and pulse constraints are imposed within a
unified differentiable objective.  A central ingredient is that the total
duration is also optimized as a time scale of the same process, rather than
selected by an external scan.  Control, evolution, and time are therefore
not separated into independent subproblems; they are organized as different
aspects of one controlled dynamical object.  The learned object is therefore
not only a waveform to be propagated, but a controlled physical process
whose internal structure and local limitations can be read out within the
same differentiable representation.

We test this idea in two settings: ordinary rotation gates, where the target
is specified by the final operation, and a geometric gate, where the path
itself is part of the operation.  For rotation gates driven by the available
$\sigma_x$ and $\sigma_z$ controls, the representation is given only the
Bloch dynamics, the gate requirement, amplitude bounds, and a time cost.
The learned process recovers the organization expected for bounded
single-qubit control: the optimized time scale, the saturated-pulse tendency,
and the division of work between the two control channels follow the target
rotation geometry.  This provides a nontrivial validation of the
representation: although the optimizer is implemented by a neural network,
the learned controlled evolution is organized by the underlying control
physics rather than by an arbitrary waveform fit.  The same gate family then
tests what survives under additional physical requirements: smooth,
hardware-friendly pulses pay an extra duration cost while preserving the
geometric organization of the control process.

We then use a geometric $Z$ gate to move from final-operation control to a
task where correctness depends on the path.  Geometric phases can depend on
the path rather than only on the final state and can reduce sensitivity to
certain control errors
\cite{Berry1984GeometricPhase,Aharonov1987CyclicPhase,Wilczek1984GaugeStructure,Zanardi1999HolonomicQC,Zhu2002NonadiabaticGeometric,Sjoqvist2012NonadiabaticHolonomic,Zhang2023GeometricHolonomicReview};
experiments in NMR, superconducting circuits, and solid-state spins show that
they are also concrete control tasks
\cite{Jones2000GeometricNMR,Feng2013NHQCExperiment,Abdumalikov2013NonAbelian,Nagata2018HolonomicSpin}.
For such a gate, the reference trajectory must satisfy a local relation with
the control direction so that the accumulated phase is geometric rather than
dynamical.  Because the fields and reference trajectory are generated within
the same differentiable process, the optimizer can coordinate the two control
channels with the path rather than assign this coordination afterward.  This
path-level requirement therefore gives a stringent test of whether the
representation can organize constraints that live along the evolution, not
only at the final time.

The representation also makes failures of this coordination visible.  In our
learned solutions the residual geometric error concentrates where the control
direction turns most rapidly and the field--trajectory relation is hardest to
maintain.  This information can be fed back into the loss by giving the
difficult intervals greater weight during training.  The resulting controls
realize a phase-closed geometric $Z$ gate with negligible dynamical phase and
reduced residual geometric error, showing that path structure can be read out
and used to refine the control process.  In a matched piecewise-constant
reference, the same local weighting more readily shifts the error to other
parts of the path, whereas the continuous representation can absorb the
correction by readjusting the fields and trajectories together.

The central claim is therefore representational: quantum optimal control can
be treated not only as pulse optimization, but as the design of a controlled
dynamical process in which fields, trajectories, constraints, and time are
aspects of one physical evolution
\cite{Peirce1988QuantumOCT,Brif2010QuantumControlReview,Koch2022QuantumTechnologies}.
Their separation in standard formulations is a limitation of representation,
not a requirement of the physics.  In the formulation developed here, the
optimized object is a physical process whose internal organization can be
read out, checked, and modified.  The point is not only to solve the control
problem, but to see through it: to expose how an optimal controlled evolution
is physically organized and make that organization usable for refinement.

\section{Methods}

We formulate single-qubit gate design in the Bloch-vector representation,
where a quantum state is described by $\rr=(x,y,z)^T$.  The qubit is driven
by two tunable controls through
$H_c(t)=[\Omega(t)\sigma_x+\Delta(t)\sigma_z]/2$.  These controls generate a
time-dependent vector field on the Bloch ball, and the controlled evolution
is written compactly as
\begin{equation}
    \dot{\rr}=F[\rr,\Omega,\Delta].
    \label{eq:bloch_main}
\end{equation}
For the rotation-gate calculations, $F$ is the dissipative Bloch vector
field with longitudinal relaxation and transverse dephasing,
$F=(-\Delta y-\Gamma_2x,\,\Delta x-\Omega z-\Gamma_2y,\,
\Omega y-\Gamma_1z+\gamma_\downarrow-\gamma_\uparrow)^T$.  The derivation,
the definitions of the rates, and the numerical values used in each
calculation are given in \appsecref{supp:physical_model}.

The target gate is defined by the ideal action that the final evolution
should realize on a set of input Bloch states.  For the rotation gates
studied below, this target is the corresponding unitary rotation of the
Bloch sphere.  Dissipation is not part of the desired operation itself; it
belongs to the physical implementation through which the controls must
realize the gate.  The optimization problem is therefore to choose
$\Omega(t)$, $\Delta(t)$, and the duration $T$ so that the implemented
evolution matches the ideal target as closely as possible under the dynamics
in Eq.~\eqref{eq:bloch_main}.

\begin{figure*}[t]
    \centering
    \includegraphics[width=0.9\textwidth]{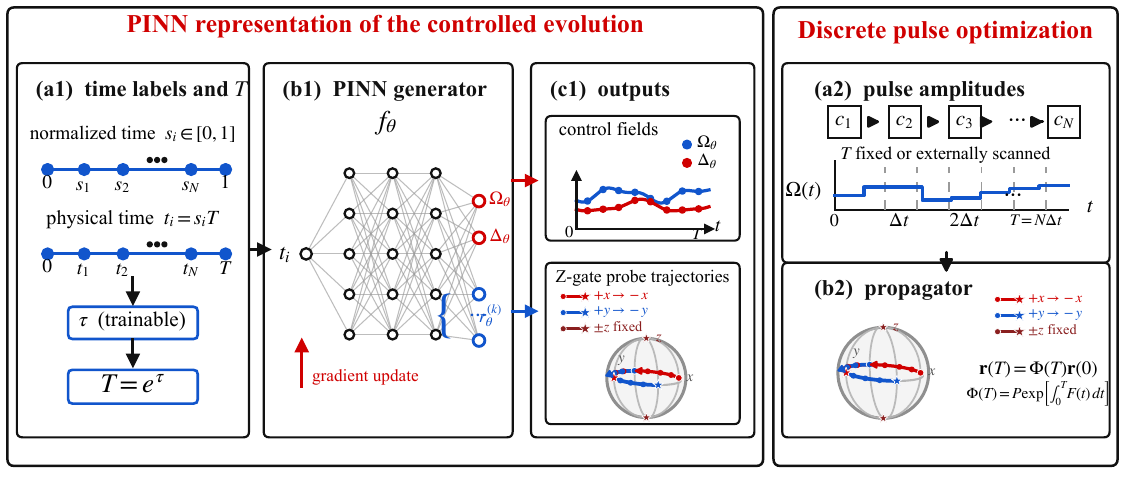}
    \caption{Discrete-pulse optimization and a PINN representation of controlled gate dynamics.  \emph{Left: PINN representation.}  \panlabel{fig.method_overview}{a1}(a1) Sample labels are mapped to physical times through the trainable duration $T=e^\tau$.  \panlabel{fig.method_overview}{b1}(b1) A physics-informed neural network (PINN) maps $t_i$ to controls and probe trajectories.  \panlabel{fig.method_overview}{c1}(c1) The outputs are differentiable over the gate interval; the Bloch sphere sketches representative $Z$-gate probe motion.  The learned fields are checked after training by independent fourth-order Runge--Kutta (RK4) propagation.  \emph{Right: discrete-pulse reference.}  \panlabel{fig.method_overview}{a2}(a2) The optimized object is a list of pulse amplitudes on a fixed or externally scanned time grid.  \panlabel{fig.method_overview}{b2}(b2) State motion is obtained by propagating the pulse, and the target operation is evaluated at the final time.}
    \label{fig:method_overview}
\end{figure*}

\subsection{Controlled dynamics and gate reconstruction}

We evaluate the implemented operation using probe states along $+\bm e_z$, $-\bm e_z$, $+\bm e_x$, and $+\bm e_y$, which are sufficient to reconstruct the affine Bloch map of a one-qubit channel.  Figure~\ref{fig:method_overview} summarizes the two formulations compared below.

The standard discrete-pulse formulation fixes a framework and then optimizes inside it~\cite{Khaneja2005GRAPE,Glaser2015TrainingCat,Koch2022QuantumTechnologies}.  The total time $T$ is chosen externally, the interval $[0,T]$ is divided into $N$ bins, and the control fields are represented by bin amplitudes $c_1,\ldots,c_N$ [Fig.~\ref{fig:method_overview}\panref{fig.method_overview}{a2}].  For a chosen pulse, the state trajectories are obtained by propagating Eq.~\eqref{eq:bloch_main} under the piecewise-constant fields [Fig.~\ref{fig:method_overview}\panref{fig.method_overview}{b2}], and the gate is checked at the final time through a terminal loss.  The dynamics are therefore enforced by propagation, while changing $T$ requires a separate optimization or scan.

The gate objective is expressed at the level of the whole process.  A one-qubit open-system evolution implements an affine map $\rout=M\rin+\bm c$ on the Bloch ball.  Because the four probes are chosen along coordinate directions, the implemented channel $(M,\bm c)$ is read directly from their final states and compared with the target channel $(\Mtar,\ctar)$ through
\begin{equation}
    \Lgate=\lambda_M\|M-\Mtar\|_F^2+
    \lambda_c\|\bm c-\ctar\|_2^2,
    \label{eq:Lgate_main}
\end{equation}
with $\ctar=\bm 0$ for the ideal target rotations used here; the explicit target matrices are given in \appsecref{supp:physical_model}.

\subsection{Physics-informed representation and optimization}

The PINN-based representation uses a different optimized object.  Instead of the discrete pulse amplitudes, we introduce a continuous generator with parameters $\theta$ that is queried by the physical time $t$, as shown in Fig.~\ref{fig:method_overview}\panref{fig.method_overview}{b1}.  Its output is a vector of heads that defines the two control fields and the four probe-trajectory functions at the same time.

The first two output heads are mapped to bounded control fields,
\begin{align}
    \Omega(t)&=\Omega_{\max}\tanh u_\Omega(t),\\
    \Delta(t)&=\Delta_{\max}\tanh u_\Delta(t),
    \label{eq:control_param_main}
\end{align}
while the remaining heads define the four probe trajectories introduced above,
\begin{equation}
    \rr^{(k)}(t)=\rr_{\rm in}^{(k)}+g(t)\bm u^{(k)}(t),
    \qquad g(0)=0 ,
    \label{eq:traj_param_main}
\end{equation}
so that the initial condition is built into the representation; representative generated trajectories are illustrated in Fig.~\ref{fig:method_overview}\panref{fig.method_overview}{c1}.  The sampled time points used in training are checkpoints at which these generated functions are tested against physics, not objects being memorized.  The important point is that the controls and the auxiliary trajectory functions are produced by the same parameter set and constrained by the same Bloch equation.  Updating $\theta$ therefore changes the fields and the generated probe paths within one differentiable representation.  The network architecture, output heads, and trajectory parametrization are detailed in \appsecref{supp:pinn}.

A joint representation of fields and trajectories is useful only if the two are physically consistent.  We impose that consistency by bringing the Bloch equation into the optimization itself.  The mechanism is a feedback loop: at sampled physical times $t_i$, we read out how far the generated trajectory deviates from the Bloch equation, and use that deviation to drive the generator back toward consistency.

Concretely, the deviation, or residual, compares the time derivative of the generated trajectory against the field it should satisfy:
\begin{equation}
    \bm R^{(k)}(t_i)=
    \dot{\rr}^{(k)}(t_i)
    -F[\rr^{(k)}(t_i),\Omega(t_i),\Delta(t_i)] ,
    \label{eq:dyn_residual_main}
\end{equation}
The corresponding dynamical loss penalizes this residual along the whole path:
\begin{equation}
    \Ldyn=\frac{1}{4N_s}\sum_{k=1}^{4}\sum_{i=1}^{N_s}
    \left\|\bm R^{(k)}(t_i)\right\|_2^2 ,
    \label{eq:Ldyn_main}
\end{equation}
This term closes the loop by sending the equation error back through the same generator.  Because both the trajectory functions and the fields come from the same generator, the residual's gradient flows back into the same parameter set $\theta$.  The trajectory functions are therefore not used as independent physical degrees of freedom; they are auxiliary trainable functions constrained by the Bloch equation, so their role is to make dynamical consistency visible and differentiable during optimization.

What closes the loop is differentiability: because the network is queried directly by the physical time $t$, automatic differentiation supplies $\dd\rr^{(k)}/\dd t$, and the residual is read out at every sampled instant without any separate propagation.  Sampling points and residual weights are given in \appsecref{supp:pinn}.

Contrast this with the propagator-based formulation of Fig.~\ref{fig:method_overview}\panref{fig.method_overview}{b2}: there the Bloch equation is solved by forward propagation, the equation error is not an explicit training signal, and the optimizer receives the final-state error from the propagated trajectory.

Once the controlled process has been represented as continuous functions, the total duration becomes an emergent quantity rather than a part of the framework fixed in advance, as shown in Fig.~\ref{fig:method_overview}\panref{fig.method_overview}{a1}.  In the discrete-pulse representation the duration is fixed first: changing it changes the time grid on which the pulse amplitudes are defined, so the time is an external hyperparameter and can only be tuned through a separate scan.  In the continuous representation the fields and trajectories are functions of physical time, and the duration is the physical time scale over which they must meet the gate; the optimization is free to settle it by stretching or compressing the interval on which the same process is queried.

We implement this by introducing an unconstrained scalar $\tau$ and defining
\begin{equation}
    T=e^\tau .
    \label{eq:T_tau_main}
\end{equation}
The trainable variables are then $(\theta,\tau)$ rather than $\theta$ alone: $\theta$ defines the continuous fields and trajectories, while $\tau$ defines the physical length of the interval on which these functions are evaluated.  To evaluate the process during training, we keep a fixed normalized set of labels $s_i\in[0,1]$ and map them to physical query points through the learnable duration,
\begin{equation}
    t_i=s_iT=s_i e^\tau ,
    \label{eq:ti_siT_main}
\end{equation}
exactly the rescaling shown in Fig.~\ref{fig:method_overview}\panref{fig.method_overview}{a1}.  The actual input to the network is the physical time $t_i$:
\begin{equation}
    f_\theta(t_i)\longrightarrow
    \{\Omega(t_i),\Delta(t_i),\rr^{(1)}(t_i),\ldots,\rr^{(4)}(t_i)\} .
    \label{eq:network_query_time_main}
\end{equation}
The duration also carries a cost.  With the dynamical loss $\Ldyn$ of Eq.~\eqref{eq:Ldyn_main} and the gate loss $\Lgate$ of Eq.~\eqref{eq:Lgate_main} now in place, we add a time penalty
\begin{equation}
    \LT=T .
    \label{eq:LT_main}
\end{equation}
All three terms are minimized together: among processes that respect the Bloch dynamics and meet the gate, this term biases the optimizer toward the shorter compatible duration.  The duration is thus not a post-processing parameter added after a fixed-time pulse has been found; it is a time-scale degree of freedom of the continuous controlled process itself.  How $\tau$ enters the network queries and the gradient computation is detailed in \appsecref{supp:pinn}.

\subsection{Loss functions and training protocol}

The total objective gathers the requirements above,
\begin{equation}
    \Ltotal(\theta,\tau)=
    \alpha_{\rm dyn}\Ldyn+
    \beta_{\rm gate}\Lgate+
    \lambda_T\LT+
    \Lreg,
    \label{eq:Ltotal_main}
\end{equation}
and is minimized over $(\theta,\tau)$ within one optimization loop.  In the discrete-pulse formulation the dynamics enter through forward propagation of a chosen pulse, whereas here the equation error, the gate objective, the time cost, and any pulse-shape penalties are coupled inside one differentiable objective.  When pulse-shape regularization is used we take
\begin{equation}
    \Lreg=
    \chi_{\rm amp}\mathcal L_{\rm amp}+
    \zeta_{\rm smooth}\mathcal L_{\rm smooth}+
    \eta_{\rm bnd}\mathcal L_{\rm boundary},
    \label{eq:Lreg_main}
\end{equation}
whose three terms penalize amplitude, smoothness, and boundary behavior respectively.  The regularized, the smooth-envelope, and the no-regularization (minimal) cases differ only in which of these terms are active.  The explicit forms and the per-prescription weights are given in \appsecref{supp:pinn}.

The construction above optimizes a controlled process against requirements imposed on the final state.  Some operations, however, place their requirements on the path itself rather than only on the endpoint.  The same representation accommodates such path-defined requirements without changing the optimized object: the generator keeps producing the fields and the trajectories, and additional differentiable terms impose the path constraints along the process.  How these constraints enter, how they interact with the control direction along the path, and what they expose about the two representations are the subject of the geometric-gate results in Sec.~\ref{sec:geometric_gate}.

\subsection{Independent validation}

Because the Bloch equation is placed inside the optimization loop, the learned process must be checked independently rather than trusted from the training loss.  After training we keep only the learned controls $\Omega(t),\Delta(t)$ and the learned duration $T$, and we propagate the Bloch equation with an external fourth-order Runge--Kutta (RK4) solver.  The independently propagated final states reconstruct the implemented channel, and we evaluate it through the process fidelity
\begin{equation}
    F_{\rm proc}=\operatorname{Tr}(\chi_{\rm tar}\chi_{\rm imp})
    \label{eq:Fproc_main}
\end{equation}
written in the unit-trace Choi convention, equivalently a trace-orthonormal Pauli process convention.  This independent propagation ensures that any structure reported in the Results is a property of a real controlled evolution, not of the training loss.  The channel-fidelity convention and the RK4 validation details are provided in \appsecref{supp:validation}.

\section{Results}

We now apply the continuous physics-informed representation to two
single-qubit gate-design problems.  The first is a family of ordinary
rotation gates, for which correctness is specified by the final operation.
This endpoint-defined setting allows us to examine how a bounded-control
solution is organized when the duration and the fields are optimized within
the same differentiable process.  The second is a geometric gate, for which
correctness also depends on how the state moves along the path.  This
path-defined setting allows us to examine whether local geometric
constraints can be represented, diagnosed, and refined within the same
framework.

\subsection{Rotation gates: emergent control geometry}

\begin{figure}[t]
  \centering
  \includegraphics[width=0.9\columnwidth]{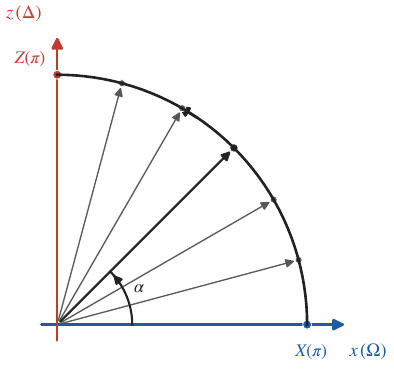}
  \caption{Rotation axes in the controlled $x$--$z$ plane.  The family $\hat n(\alpha)=(\cos\alpha,0,\sin\alpha)$ interpolates between $X(\pi)$ and $Z(\pi)$ using the available $\Omega$ and $\Delta$ channels.}
  \label{fig:rotation_axes}
\end{figure}

The rotation-gate family is used here not merely as a fidelity benchmark,
but as a test of whether the learned process organizes itself according to
the physics of the available controls.  With only the $\sigma_x$ and
$\sigma_z$ controls available, the directly generated family consists of
rotations whose axes lie in the $x$--$z$ plane,
\begin{equation}
    \hat n(\alpha)=(\cos\alpha,0,\sin\alpha),
    \qquad
    \alpha\in[0,\pi/2],
\end{equation}
shown in Fig.~\ref{fig:rotation_axes}.  The angle $\alpha$ parameterizes
the rotation-axis direction, from $\alpha=0$ ($X$ gate) to
$\alpha=\pi/2$ ($Z$ gate), so that both control channels enter along the
family.  All results below are evaluated by independent RK4 propagation and
channel reconstruction after training; the network-generated trajectories
are used only during the physics-informed optimization.  As a visual guide
to the controlled processes analyzed in this section, Movie S1 of the
Supplemental Material~\cite{supplemental} shows time-resolved Bloch
trajectories synchronized with the control fields for representative
rotation-gate examples.

\resultsubpart{Minimal prescription}

We first use a minimal prescription to expose the intrinsic organization:
$\Omega_{\max}=\Delta_{\max}=8.0$, initial duration $T_0=0.7$,
time-cost coefficient $\lambda_T=1$, and no smoothness, amplitude,
boundary, or envelope regularization.  Numerical weights and sampled-time
settings are collected in \appsecref{supp:pinn}.  In this setting the
optimizer is given only the equation of motion, the gate objective, the
amplitude bounds, and the time cost.  No pulse ansatz, channel-division
rule, or reference time scale is supplied.

\begin{figure}[t]
  \centering
  \includegraphics[width=0.9\columnwidth]{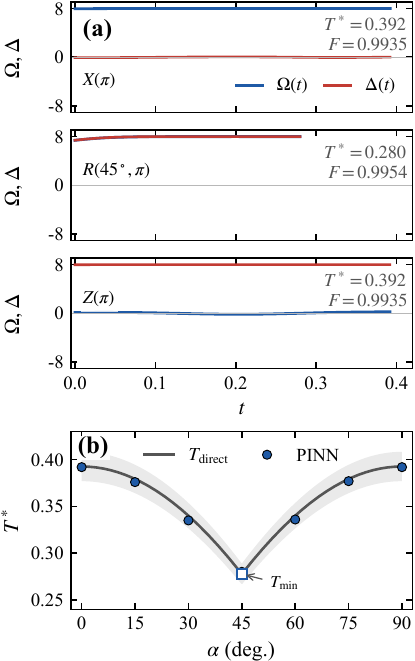}
  \caption{Organization under the minimal prescription.  (a) Representative controls for $X(\pi)$, $R(45^\circ,\pi)$, and $Z(\pi)$ recover saturated square-pulse structure without prescribing a square-pulse ansatz.  (b) The optimized durations $T^*$ for all seven in-plane gates follow the Pontryagin-based direct-control prediction $T_{\rm direct}$ of Eq.~\eqref{eq:T_direct_results}; the shaded band marks $\pm4\%$.}
  \label{fig:rotation_minimal}
\end{figure}

Figure~\ref{fig:rotation_minimal}(a) shows the learned solutions for seven
in-plane gates, all reaching average gate fidelities above $99.3\%$ under
independent RK4 validation.  The pulse shapes are not arbitrary: the
$X(\pi)$ solution uses an almost saturated $\Omega$ channel with
$\Delta\simeq0$, the $Z(\pi)$ solution exchanges the two roles, and the
near-$45^\circ$ rotation uses both bounded channels simultaneously.  A
saturated square pulse is the form one would impose by hand for a
bounded-control problem, yet here no square-pulse ansatz was supplied; it
emerges from the optimization.

The strongest evidence for physical organization is the time scale.
Figure~\ref{fig:rotation_minimal}(b) shows that the learned durations
$T^*(\alpha)$ form a U-shaped curve: shortest near $\alpha=45^\circ$,
where the two channels share the rotation most efficiently, and longer at
the endpoints, where a single channel must carry the full rotation.  The
curve matches a simple analytical reference.  Pontryagin's maximum
principle identifies the saturated-control extremals of the corresponding
bounded-control problem, for which the control vector $(\Omega,0,\Delta)$
aligns with the desired axis $(\cos\alpha,0,\sin\alpha)$
\cite{Boscain2006TimeMinimal,Garon2013TimeOptimalSU2,Boozer2012SU2Synthesis,Dionis2023PiecewiseTimeOptimal};
with independent bounds $|\Omega|,|\Delta|\leq\Omega_{\max}$ at least one
channel saturates, giving the direct-control reference scale
\begin{equation}
    T_{\rm direct}(\alpha)
    =
    \frac{\varphi\,\max(\cos\alpha,\sin\alpha)}{\Omega_{\max}} .
    \label{eq:T_direct_results}
\end{equation}
The optimized durations follow this reference to within a few percent
[Fig.~\ref{fig:rotation_minimal}(b)].

What makes this agreement nontrivial is that the optimizer never sees the
Pontryagin result.  It is not asked to imitate a square pulse, to divide the
work between channels according to a prescribed rule, or to match a
reference-time formula.  The direct-control time scale is recovered as a
structural property of the optimized process.  Thus, in the endpoint-defined
setting, the neural representation does not only find successful
waveforms; it recovers a controlled evolution whose duration, pulse
behavior, and channel use can be read out as physical structure.

\resultsubpart{Pulse constraints}

\begin{figure}[t]
  \centering
  \includegraphics[width=0.9\columnwidth]{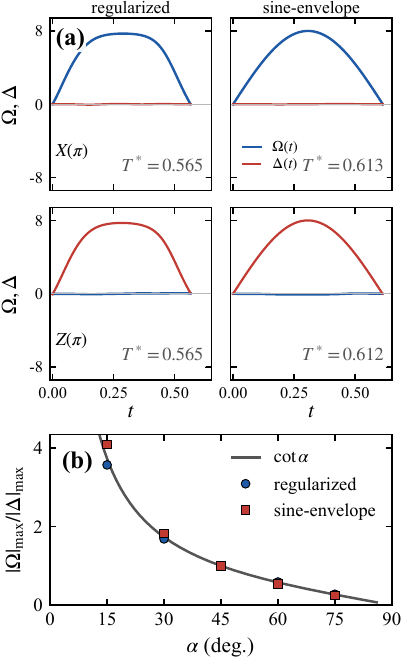}
  \caption{Pulse-shape prescriptions and the geometric channel-division law.  (a) Regularized and sine-envelope controls for the in-plane $X(\pi)$ and $Z(\pi)$ gates, with $T^*$ marked in each panel.  (b) The peak-amplitude ratio $|\Omega|_{\max}/|\Delta|_{\max}$ follows $\cot\alpha$ for both prescriptions, showing that the channel-division law is set by the target geometry and survives pulse-shape constraints.  Numerical weights are listed in \appsecref{supp:pinn}.}
  \label{fig:pulse_comparison}
\end{figure}

The minimal solutions expose the intrinsic time law, but their square-like
form is not the pulse shape preferred by hardware.  We therefore repeat the
optimization under two smoother prescriptions within the same continuous
representation.  The first is a regularized prescription with very weak
amplitude, smoothness, and boundary penalties
($\chi_{\rm amp}=\zeta_{\rm smooth}=10^{-4}$, $\eta_{\rm bnd}=10^{-3}$);
Fig.~\ref{fig:pulse_comparison}(a) shows
that even these tiny penalties already produce pulses that are smooth and
well behaved at the gate endpoints.  The second is a sine-envelope
prescription, in which both controls are multiplied by $\sin(\pi t/T)$ so
that they vanish exactly at the gate endpoints.  The penalty weights and
the envelope form are given in \appsecref{supp:pinn}.

For the $X(\pi)$ and $Z(\pi)$ gates the sine-envelope pulse takes a longer
optimized duration $T^*$ than the regularized one, as marked in each panel.
The regularized pulse is only softly penalized at the endpoints and can stay
near its plateau across the interval, whereas the sine-envelope is forced
to ramp up and back down, so for the same amplitude bound it accumulates
less rotation per unit time.  What does not change across the two
prescriptions is the division of work between the two control channels.

If the target axis lies in the $x$--$z$ plane and the control vector follows
that axis, then the peak amplitudes of the two channels satisfy
\begin{equation}
    \frac{|\Omega|}{|\Delta|}
    \simeq
    \frac{\cos\alpha}{\sin\alpha}
    =
    \cot\alpha .
    \label{eq:division_law}
\end{equation}
In both the regularized and the sine-envelope solutions the peak-amplitude
ratio follows $|\Omega|_{\max}/|\Delta|_{\max}\simeq\cot\alpha$ across the
family [Fig.~\ref{fig:pulse_comparison}(b)].  Changing the pulse-shape
constraint, smoothing the pulse or forcing it to vanish at the endpoints,
changes the pulse shape and the optimized duration but not the division of
work.  The channel-division law is set by the target geometry and preserved
because the fields and trajectories remain coupled by the Bloch equation
inside the optimization loop.

A dissipative ablation (\appsecref{supp:physical_model}) gives a limited
check on the role of open-system terms.  For the rates tested here,
including dissipation in the Bloch-equation loss slightly changes the
detailed pulse shape but does not systematically shorten the optimized time
or improve the validated fidelity.  The limiting effect is the irreversible
shrinkage of the Bloch vector, which unitary controls cannot undo.

The rotation-gate results establish the first role of the representation:
for endpoint-defined gates, the learned process can recover physically
expected control organization without being told that organization in
advance.  We now turn to a geometric gate, where the requirement is
stronger.  The trajectory itself must satisfy a local condition along the
path, so the relation between the fields and the trajectory becomes part of
the gate requirement rather than a side effect of the final operation.

\FloatBarrier

\subsection{Geometric gates: coordination and refinement}
\label{sec:geometric_gate}

\begin{figure}[t]
  \centering
  \includegraphics[width=0.9\columnwidth]{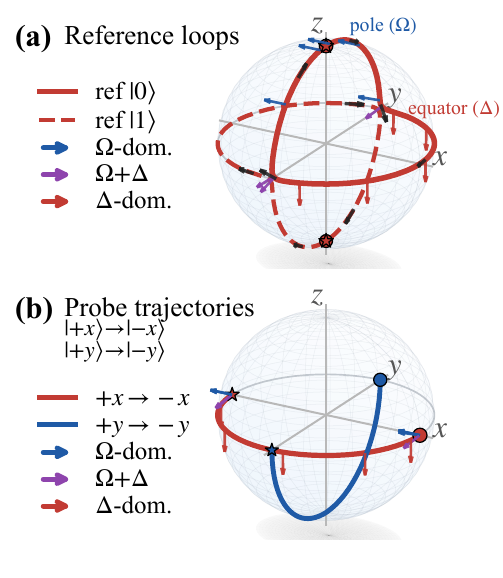}
  \caption{Path structure of the geometric $Z$ gate.  (a) Reference states follow closed loops that carry the geometric phase.  Arrows show the control-field vector $\bm h(t)=(\Omega,0,\Delta)$, drawn perpendicular to the local reference direction.  The colors denote the dominant channel: blue for $\Omega$ dominance, purple for coexistence, and red for $\Delta$ dominance.  (b) Probe states reconstruct the target $Z$ rotation.  The $+z$ state is fixed, so $|{+}x\rangle\rightarrow|{-}x\rangle$ and $|{+}y\rangle\rightarrow|{-}y\rangle$ are the visible transitions.}
  \label{fig:geometric_paths}
\end{figure}

A geometric $Z$ gate has the same terminal action as the rotations above:
it flips the transverse Bloch components while leaving the $z$ component
unchanged.  Here it is realized by the geometric phase acquired around a
closed loop rather than by accumulated dynamical phase.  The phase depends
on the solid angle enclosed by the loop, not on how fast it is traversed.
Figure~\ref{fig:geometric_paths} shows the setup.  Two reference states
traverse closed loops that carry this phase
[Fig.~\ref{fig:geometric_paths}(a)], while probe states are acted on by the
resulting gate [Fig.~\ref{fig:geometric_paths}(b)].  Movie S2 of the
Supplemental Material~\cite{supplemental} provides the corresponding
time-resolved visualization of the optimized geometric $Z$-gate path,
including the synchronized control fields and the instantaneous control
direction.

For the operation to be geometric, the reference states must satisfy three
conditions throughout the path, not only at the final time.  Two orthogonal
reference states, $\hat{\bm n}_+(t)$ and $\hat{\bm n}_-(t)$, serve as the
cyclic states that carry the geometric phase.  They are the $Z$-gate
eigenstates $\pm\hat z$, chosen so that each loops away from and back to
its starting direction at $t=T$.  The geometric-phase difference accumulated
around the two loops realizes the gate.  First, each loop closes.  Every
reference state returns to its initial ray at $t=T$, so that the loop
encloses a well-defined solid angle.  Second, the two references stay
antipodal, $\hat{\bm n}_+(t)\cdot\hat{\bm n}_-(t)=-1$, so they span a
complete basis at each instant.  Third, the local geometricity condition
\begin{equation}
    E(t)=\langle\phi(t)|H(t)|\phi(t)\rangle=\tfrac12\,\bm h(t)\cdot\hat{\bm n}(t)=0,
    \quad 0\le t\le T,
    \label{eq:geometricity_results}
\end{equation}
must hold at every instant, where $\bm h(t)=(\Omega(t),0,\Delta(t))$ is the
instantaneous control-field vector.  Equation~\eqref{eq:geometricity_results}
removes the dynamical phase pointwise.  It requires the control-field
direction $\bm h(t)$ to stay perpendicular to the reference direction
$\hat{\bm n}(t)$ throughout.  Its component form
$E=\tfrac12(\Omega n_x+\Delta n_z)=0$ exposes the structure behind this
perpendicularity, which the arrows in Fig.~\ref{fig:geometric_paths}(a)
visualize.  Each arrow marks $\bm h(t)$ at a point of the loop and is drawn
perpendicular to the local reference direction, as
Eq.~\eqref{eq:geometricity_results} requires.  The arrows are colored by
which channel carries the motion.  Blue marks $\Omega$-dominant intervals,
red marks $\Delta$-dominant intervals, and purple marks intervals where the
two act together.  This color convention is used in all later figures.

These path conditions are imposed as differentiable loss terms on the same
learned process used for the rotation gates, not checked only after a pulse
has been found.  The full geometric-gate objective is
\begin{equation}
\begin{aligned}
\mathcal L_{\rm total}^{\rm geom}={}&
\lambda_{\rm dyn}\Ldyn+\lambda_{\rm gate}\Lgate+\lambda_T\LT+\Lreg\\
&+\lambda_{\rm geo}\mathcal L_{\rm geo}+\lambda_{\rm cyc}\mathcal L_{\rm cyc}
+\lambda_{\rm pur}\mathcal L_{\rm pur}+\lambda_{\rm orth}\mathcal L_{\rm orth},
\label{eq:Lgeom_total}
\end{aligned}
\end{equation}
whose first line is inherited from the rotation-gate objective and whose
second line carries the path conditions specific to the geometric gate.  The
central added term is the geometricity loss
$\mathcal L_{\rm geo}=\frac{1}{N_s}\sum_i[E_+(t_i)^2+E_-(t_i)^2]$, which
penalizes violations of Eq.~\eqref{eq:geometricity_results}.  The
cycle-closure loss $\mathcal L_{\rm cyc}$ drives each reference state back
to its initial ray at $t=T$.  The purity and orthogonality penalties keep
the two references on the Bloch sphere and antipodal.  The explicit forms
of the path losses and all per-term weights are given in
\appsecref{supp:geometry}.  The implemented rotation is reconstructed from
three Cartesian probe trajectories $+x,+y,+z$ as in the rotation-gate tasks.
For the $Z$ gate the eigenstate $+z$ is fixed, so only
$|{+}x\rangle\rightarrow|{-}x\rangle$ and
$|{+}y\rangle\rightarrow|{-}y\rangle$ are visible at the final time
[Fig.~\ref{fig:geometric_paths}(b)].  The probe trajectories verify the
final $Z$ operation, while the reference loops and the local geometricity
condition test whether the implementation suppresses dynamical phase along
the path.

\begin{figure}[t]
  \centering
  \includegraphics[width=0.9\columnwidth]{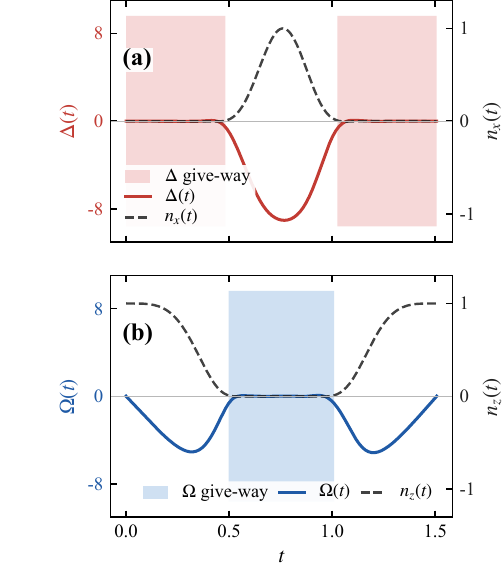}
  \caption{Automatic coordination in the geometric $Z$ gate, shown on a common time axis with the same color convention as Fig.~\ref{fig:geometric_paths}.  (a) $\Delta(t)$ (red) versus $n_x(t)$ (dashed).  The red-shaded bands mark where $|n_x|\to0$, which forces $\Delta\to0$ because $E=\tfrac12(\Omega n_x+\Delta n_z)$ then reduces to $\tfrac12\Delta n_z$.  (b) $\Omega(t)$ (blue) versus $n_z(t)$ (dashed).  The blue-shaded bands mark where $|n_z|\to0$, which forces $\Omega\to0$.  Each control channel drops exactly where its geometric coefficient drops.  The anti-correlation is the visible signature of the field--trajectory coordination.}
  \label{fig:coordination_main}
\end{figure}

The geometric-gate objective is stiffer than the rotation-gate objective
because the path equalities must be maintained along the entire trajectory.
In practice we use a larger network and a staged schedule.  The Bloch
dynamics, final gate, and cycle closure are emphasized first, and the
geometricity weight is raised only after a physically valid controlled path
has formed.  This avoids the shortcut in which the algebraic quantity
$E(t)$ is reduced on the training grid while the independently propagated
trajectory fails the gate.  With this schedule the two-control geometric
$Z$ gate reaches $F_{\rm proc}>0.999999$ at an optimized duration
$T^*\simeq1.5$ under independent RK4 validation.  The numerical weights,
architecture, and schedule are given in \appsecref{supp:geometry}.

\resultsubpart{Automatic coordination}

Meeting the geometricity condition along the whole path requires the
control fields and the reference trajectories to coordinate locally, not
only to reach the correct final state.  Figure~\ref{fig:coordination_main}
places the reference-state direction and the control fields on a common time
axis.  The same color convention as Fig.~\ref{fig:geometric_paths} is used:
red for the $\Delta$ channel and blue for the $\Omega$ channel.

The component form of Eq.~\eqref{eq:geometricity_results},
$E=\tfrac12(\Omega n_x+\Delta n_z)=0$, makes the coordination concrete.
When the reference state points along $\hat z$ (so $n_z\approx1$ and
$n_x\approx0$), the $\Omega$ term drops out and
$E\approx\tfrac12\Delta\,n_z$.  This quantity can vanish only if
$\Delta\approx0$, so the gate is then driven by $\Omega$ alone.  When the
reference state points along $\hat x$ instead ($n_x\approx1$ and
$n_z\approx0$), the roles reverse and $\Omega$ must give way.  Along a
closed path the reference state moves between these two limits.  The two
control channels therefore exchange roles.  Each channel is driven to zero
precisely where its geometric coefficient vanishes and recovers where it no
longer does.

The optimization recovers this rule without being told it.
Figure~\ref{fig:coordination_main} shows the learned controls obeying the
inference along the whole path, not only at the poles and the equator where
it is exact.  Each channel drops to zero where its geometric coefficient
vanishes and recovers where it no longer does.  This synchronized motion is
not imposed by the parametrization.  It emerges from optimizing the fields
and the reference-trajectory functions under the same dynamical and
geometric constraints.  A pulse-only formulation can still search over
pulses, but it does not make this field--trajectory coordination a native
optimization variable.  In the continuous representation, by contrast, the
coordination is part of the learned physical process itself.

\resultsubpart{Turning bottleneck}

\begin{figure}[t]
  \centering
  \includegraphics[width=0.9\columnwidth]{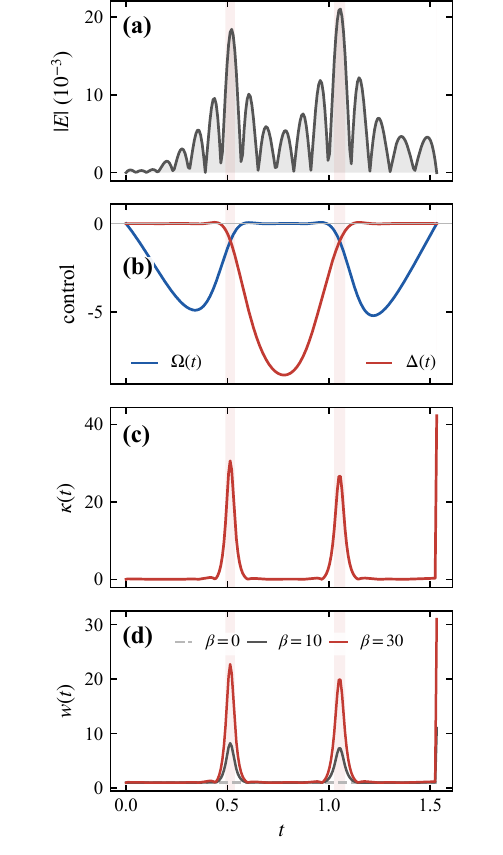}
  \caption{Local geometricity error in the geometric $Z$ gate.  (a) Pointwise geometricity error $|E(t)|$ for the baseline solution, peaking at two localized intervals.  (b) Control fields $\Omega(t)$ and $\Delta(t)$.  The error peaks coincide with moments where the control direction reorients rapidly.  (c) Turning-rate diagnostic $\kappa(t)$, peaking at the same intervals as $|E(t)|$.  (d) Turning-weight factor $w(t)=1+\beta\kappa_{\rm norm}(t)$ for three values of $\beta$.  As $\beta$ grows, $w(t)$ concentrates onto the rapidly turning intervals and converts the local diagnosis into position-dependent design pressure.}
  \label{fig:turning_bottleneck}
\end{figure}

Even with good coordination the geometricity error is not uniform along the
path.  Figure~\ref{fig:turning_bottleneck}(a) shows the pointwise error
$|E(t)|$ for the baseline solution.  It is small along most of the path,
rises sharply at two localized intervals, and is otherwise negligible.
To see what is special about those intervals, we turn to the control fields
themselves [Fig.~\ref{fig:turning_bottleneck}(b)].  The error peaks coincide
with moments where $\Omega(t)$ and $\Delta(t)$ reorient most rapidly.  They
therefore occur where the control direction turns sharply rather than where
the control amplitude is large.

This observation motivates a quantitative measure of how fast the control
direction turns.  We define the turning rate $\kappa(t)$ of the control
vector in the $\Omega$--$\Delta$ plane.  The explicit form and its
regularization are given in \appsecref{supp:geometry}.  A large
$\kappa(t)$ marks an interval where the control direction turns rapidly,
independently of the control amplitude.  Since $\Omega(t)$ and $\Delta(t)$
are differentiable network outputs, $\kappa(t)$ is evaluated directly by
automatic differentiation.  Figure~\ref{fig:turning_bottleneck}(c) shows
that $\kappa(t)$ peaks at the same intervals where $|E(t)|$ is large.  The
geometricity error is therefore concentrated where the control direction
turns fastest and where the reference trajectory must follow it most
quickly to keep the two perpendicular.

This diagnosis points to a design handle.  Because the geometricity error
lives at identifiable positions, the objective can penalize those positions
more heavily.  The optimizer is then asked to maintain the perpendicular
relation more carefully exactly where it is hardest.  Figure~\ref{fig:turning_bottleneck}(d)
shows the resulting turning-weight factor
$w(t)=1+\beta\kappa_{\rm norm}(t)$.  For $\beta=0$ it is uniform.  As
$\beta$ grows, it concentrates onto the rapidly turning intervals identified
above.  The local diagnosis is thereby converted into a position-dependent
design pressure.

\resultsubpart{Turning-weighted refinement}

We add the position-dependent turning weight to the geometricity loss,
leaving the scalar coefficient $\lambda_{\rm geo}$ unchanged.  The
geometricity term $\lambda_{\rm geo}\mathcal L_{\rm geo}$ in
Eq.~\eqref{eq:Lgeom_total} becomes
$\lambda_{\rm geo}\mathcal L_{\rm geo}^{\rm turn}$, with
\begin{equation}
    \mathcal L_{\rm geo}^{\rm turn}
    =
    \frac{1}{N_s}\sum_i
    w(t_i)
    \left[E_+(t_i)^2+E_-(t_i)^2\right],
    \label{eq:turn_weighted_results}
\end{equation}
where $w(t)=1+\beta\,\kappa_{\rm norm}(t)$, shown in
Fig.~\ref{fig:turning_bottleneck}(d), is the only new ingredient.
The scalar coefficient $\lambda_{\rm geo}$ keeps its staged value
($10$ in the first stage and $300$ in the second).  The factor $w(t)$
additionally up-weights the rapidly turning intervals.  For $\beta=0$,
$w(t)=1$ and $\mathcal L_{\rm geo}^{\rm turn}$ reduces to the uniform
$\mathcal L_{\rm geo}$.  Here $\kappa_{\rm norm}$ is normalized and used as
a fixed factor rather than back-propagated.  The prescription and
representative values are given in \appsecref{supp:geometry}.  Larger
$\beta$ concentrates the penalty more strongly onto the turning intervals.

\begin{figure}[t]
  \centering
  \includegraphics[width=0.9\columnwidth]{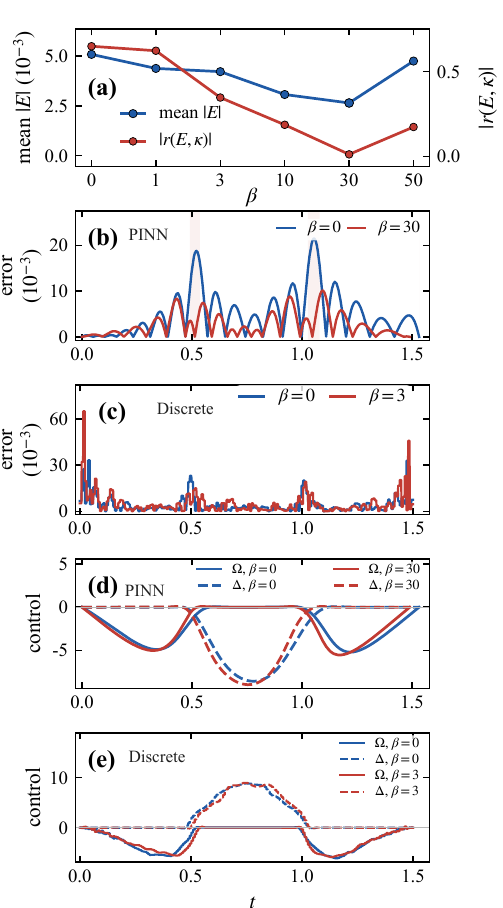}
  \caption{Turning-aware weighting of the geometricity loss.  (a) Continuous-representation scan over the weight $\beta$.  The mean pointwise geometricity error and its correlation with $\kappa(t)$ are shown, with the optimum near $\beta\approx 30$.  (b) RK4-evaluated geometricity error in the continuous representation before and after weighting the turning intervals.  The error drops without rising elsewhere.  (c) The same weighting applied to the piecewise-constant discrete baseline at the matched duration $T=1.5$.  The error drops at the targeted intervals but reappears elsewhere.  (d) Continuous controls for the two weights in (b).  (e) Discrete controls for the two weights in (c).}
  \label{fig:turn_weighted_loss}
\end{figure}

Figure~\ref{fig:turn_weighted_loss}(a) scans $\beta$ in the continuous
representation.  As $\beta$ grows from zero, the mean geometricity error
decreases and its correlation with $\kappa$ weakens.  This happens because
the optimizer is pressed harder precisely where the error was previously
largest.  The process fidelity remains above $0.999999$ throughout.  An
optimum is reached near $\beta\approx 30$.  At this value the
turning-interval error is largely eliminated while the gate accuracy and
smoothness are preserved.  Increasing $\beta$ further ($\beta=50$) raises
the error again because the over-concentrated weight disrupts the balance
among geometricity, smoothness, and gate accuracy.  Panels~(b) and~(d)
confirm this directly.  After weighting, the geometricity error at the
turning intervals drops without rising elsewhere, and the controls adjust
smoothly to keep the field and the trajectory coordinated through the
turns.  The structure read out from the process has thus been fed back into
the objective.  The difficult intervals are reduced rather than merely
relocated.

The same turning-aware weight applied to the finite piecewise-constant
baseline behaves differently [Fig.~\ref{fig:turn_weighted_loss}(c,e);
\appsecref{supp:discrete_pulse}].  The
baseline already carries its largest error near the start and end of the
path, where the reference state sits close to a pole and the geometricity
condition forces the corresponding channel toward zero faster than a
piecewise-constant pulse can suppress it.  Once the weight asks for better
control at the turning intervals, the correction perturbs the trajectory
downstream.  The error then drops where it is weighted but reappears, and
sometimes grows, elsewhere along the path.  This comparison is specific to
the matched piecewise-constant baseline used here.  Smooth-basis or
state-constrained conventional optimal-control formulations may behave
differently.  In the PINN-based representation, by contrast, the auxiliary
trajectory functions can re-coordinate with the field under the same path
constraints.  A correction at one interval is therefore absorbed rather than
transmitted along the path.

Taken together, the geometric-gate results show why path-defined tasks
benefit from making trajectories part of the optimized representation.  The
geometricity error, the control direction, and the turning rate can be
evaluated along the whole process, not only through the final gate.  These
observations can then be fed back into the objective, as the turning-aware
weight does, so that what is learned about the path reshapes the next
optimization.  In this sense, coordination, diagnosis, and refinement form a
single loop within the PINN-based representation.

\FloatBarrier

\section{Conclusion and Discussion}

This work is based on a representational observation: a quantum gate is not
only a pulse waveform, but the controlled evolution generated by that
waveform.  Pulse-centered formulations remain powerful numerical tools, yet
they naturally make the waveform the primary optimized object and leave the
induced evolution to be reconstructed or interpreted afterward.  Our aim was
to ask what changes when the gate-forming evolution itself is represented as
one differentiable physical object.  In this form, the optimized result is
not merely a pulse that reaches a target, but a controlled process in which
the fields, trajectories, duration, and path-level constraints can be
examined within the same representation.

The two examples show what this change of representation makes possible.
For endpoint-defined rotation gates, the learned process recovers the
bounded-control organization expected from the underlying dynamics without a
prescribed pulse ansatz or duration scan.  For the geometric gate, where the
constraint is distributed along the path, the same representation exposes
localized geometricity errors and converts this information into targeted
refinement.  Thus the central result is not only the synthesis of
high-fidelity gates, but the conversion of quantum control from a waveform
search into a process-level design problem: the learned evolution becomes a
physical object that can be read, diagnosed, and improved.

This diagnosis-and-refinement loop also clarifies the experimental relevance
of the approach.  In a real device, the difficulty is often not only to find
a pulse with high simulated fidelity, but to adapt the control when a
specific part of the process becomes fragile because of bandwidth limits,
transfer-function distortions, calibration errors, drifts, leakage, noise,
or path-dependent requirements.  A pulse-centered workflow can certainly
incorporate such effects, but they are often added as external constraints,
diagnostics, or redesign steps around the waveform search.  In the present
formulation, by contrast, the fields, trajectories, duration, and local
path errors are available within the same differentiable process.  Once a
failure mode is localized in the learned evolution, it can be turned
directly into a term in the objective and used to guide the next
optimization.  This does not by itself solve the full experimental
calibration problem, but it provides a concrete route toward controls that
are not only high-fidelity, but also physically diagnosable and
systematically refinable.

This perspective complements standard quantum-control methods rather than
replacing them.  Mature approaches such as gradient-ascent pulse
engineering, Krotov-type algorithms, GOAT-like analytic-control methods,
and CRAB-type smooth basis parametrizations remain powerful tools for
producing high-fidelity controls in fixed-time or parametrized problems
\cite{Khaneja2005GRAPE,Palao2003KrotovUnitary,Goerz2019KrotovPackage,Brif2010QuantumControlReview,Glaser2015TrainingCat,Machnes2018GOAT,Caneva2011CRAB,Koch2022QuantumTechnologies}.
The distinction is not what conventional control can or cannot optimize in
principle, but which physical quantities are native to the formulation.  A
continuous physics-informed representation makes time scales, trajectory
structure, local equation residuals, and path-level conditions available as
differentiable quantities that can be inspected and fed back into the
design.  In this sense, the present approach is best viewed as a
process-level representation that can work alongside conventional solvers:
standard methods may efficiently produce candidate controls, while the
continuous representation can analyze, diagnose, and refine the physical
process that those controls generate.

The same point is important for physics-informed learning and AI for
science.  The result is not simply that a PINN can synthesize a
high-fidelity gate.  A small equation error during training should not be
trusted as physical correctness by itself, since soft physics constraints
are known to be sensitive to loss imbalance, shortcut solutions, and
optimization pathologies
\cite{Karniadakis2021PIML,Cuomo2022PINNReview,Wang2021GradientPathologies,Krishnapriyan2021FailureModes}.
For this reason, the learned controls are validated by independent RK4
propagation and channel reconstruction rather than by the training loss
alone.  Once validated, however, the differentiability of the representation
becomes scientifically useful: equation residuals, trajectory relations,
turning rates, and path errors are not only diagnostics of a trained
solution, but quantities that can reshape the optimization objective.  The
neural representation therefore contributes not only numerical flexibility,
but a way of making the learned physical process readable and improvable.

Several limitations remain.  First, the present study is restricted to a
single qubit, where the Bloch-vector picture and channel reconstruction
from a small set of coordinate probes are especially transparent.  Extending
the approach to multiqubit gates will require more scalable descriptions of
the implemented channel, for example operator-level losses, randomized
input states, tensor-network-inspired representations, or other compressed
descriptions of many-body dynamics.  Second, the hardware constraints
considered here are still idealized.  Real bandwidth limits, transfer
functions, calibration nonlinearities, leakage channels, and noise spectra
should eventually be incorporated into the controlled process itself.
Third, the learnable-time formulation finds an optimized duration within a
chosen parametrization, loss balance, and training protocol, and should not
be read as a proof of global time optimality.  Finally, because the
dynamics are imposed through soft equation errors, balancing dynamical
accuracy, gate fidelity, time cost, and regularization remains a practical
issue rather than an automatic consequence of the formulation.

A natural next direction is therefore a hybrid program.  Conventional
quantum-control solvers can provide efficient high-fidelity candidate
pulses, while a continuous physics-informed representation can be used to
analyze their time scales, trajectory structure, and local bottlenecks.
For rotation gates, this suggests studying anisotropic amplitude bounds,
drift terms, experimentally measured transfer functions, and calibration
feedback.  For geometric gates, it suggests extending the coordination and
turning-aware refinement to multilevel leakage, non-Abelian holonomies, and
robustness objectives that distinguish genuine geometric protection from
ordinary pulse smoothing.  More broadly, the same idea may be useful
whenever the scientifically relevant object is not only the final operation,
but the organized physical process through which the operation is formed.

In conclusion, single-qubit gate design can be viewed not only as pulse
optimization, but as the design of a controlled dynamical process.  By
representing this process continuously and enforcing the governing
equation, the PINN-based formulation holds together fields, trajectories,
duration, constraints, and path information within one learned object.  For
rotation gates, this makes the expected control organization emerge from
the physics rather than from a prescribed ansatz.  For the geometric gate,
it makes a localized path-level failure mode visible and turns that
diagnosis into refinement.  Returning to the distinction with which the
paper began, the aim is not only to solve the control problem, but to see
through it: to expose how a gate-forming evolution is physically organized
and to make that organization usable for further design.

\begin{acknowledgments}
This work was supported by the National Natural Science Foundation
of China (Grant Nos.~12275165 and 12275212), the Young Talent Support Program for
Doctoral Students of the China Association for Science and Technology
(CAST), the Natural Science Foundation of Shaanxi Province
(Grant No.~2025JC-YBQN-055), and the Youth Innovation Team of Shaanxi Universities
(Grant No.~24JP177).
\end{acknowledgments}

\section*{Data Availability}
The program codes and datasets used in the current paper can be obtained from Ref.~\cite{repo:PINN-code}.

\FloatBarrier
\appendix

\setlength{\abovedisplayskip}{4pt plus 2pt minus 2pt}
\setlength{\belowdisplayskip}{4pt plus 2pt minus 2pt}
\setlength{\abovedisplayshortskip}{2pt plus 1pt minus 1pt}
\setlength{\belowdisplayshortskip}{2pt plus 1pt minus 1pt}

\section{Physical model, dissipative settings, and target-gate definitions}
\label{supp:physical_model}

The Bloch equation used in the main text is obtained from the driven-qubit
Lindblad master equation
\begin{equation}
\begin{aligned}
\dot\rho={}&-i[H_c(t),\rho]
+\gamma_\downarrow\mathcal D[\sigma_-]\rho
+\gamma_\uparrow\mathcal D[\sigma_+]\rho
+\frac{\gamma_\phi}{2}\mathcal D[\sigma_z]\rho ,
\end{aligned}
\label{eq:supp_master_equation}
\end{equation}
where
\begin{equation}
    \mathcal D[L]\rho=L\rho L^\dagger
    -\frac12\{L^\dagger L,\rho\}.
\end{equation}
With the Hamiltonian and Bloch-vector convention of the main text, this
master equation gives the component form of Eq.~\eqref{eq:bloch_main}.  The
longitudinal and transverse decay rates are
\begin{equation}
    \Gamma_1=\gamma_\downarrow+\gamma_\uparrow,
    \qquad
    \Gamma_2=\frac{\Gamma_1}{2}+\gamma_\phi .
    \label{eq:supp_Gamma_def}
\end{equation}
This convention corresponds to the dephasing term
$(\gamma_\phi/2)\mathcal D[\sigma_z]\rho$, so that pure dephasing contributes
$\gamma_\phi$ to the transverse decay rate.

The dissipative rates enter only the implemented dynamics.  They do not
change the target gate.  Throughout the paper, the target channel is the
ideal unitary rotation, while relaxation and dephasing are treated as part
of the physical process that the learned control must model and diagnose.
All rates are given in the same dimensionless units as the control amplitudes
and the inverse gate duration.  The rotation-gate main results use
$\gamma_\downarrow=0.05$, $\gamma_\uparrow=0$, and $\gamma_\phi=0$.
The closed-system training reference used in the dissipative ablation sets all
three rates to zero.  The geometric-gate calculations are also closed-system
calculations, with $\gamma_\downarrow=\gamma_\uparrow=\gamma_\phi=0$.

For the dissipative ablation, we compare two training procedures.  In the
first, the PINN residual uses the full dissipative Bloch equation with the
rotation-gate rates above.  In the second, the PINN residual uses the
corresponding closed-system equation, obtained by setting all dissipative
rates to zero.  After training, both learned controls are validated under the
same full dissipative dynamics.  This comparison separates the effect of
including dissipation in the physics-informed loss from the unavoidable purity
loss produced by relaxation and dephasing during the implemented evolution.
In the parameter range tested here, including dissipation in the residual
produces small changes in the detailed pulse shape, such as weak deviations
from the nearly square minimal-control profiles, but leaves the optimized
duration essentially unchanged.  It also does not yield a systematic fidelity
advantage; the validated fidelity differences remain at the $10^{-4}$--$10^{-6}$
level.  This is expected:
once relaxation or dephasing has reduced the Bloch-vector length during the
evolution, coherent control can change the direction of the state but cannot
restore the purity already lost to the environment.

A target rotation acts on Bloch vectors as a $3\times3$ rotation about an
axis $\hat n$ through angle $\varphi$.  For the in-plane family used in the
main text, the axis lies in the directly controlled $x$--$z$ plane,
\begin{equation}
    \hat n=(\cos\alpha,\,0,\,\sin\alpha),
\end{equation}
with $\alpha=0$ and $\alpha=\pi/2$ giving the $X$ and $Z$ rotations.  The
target Bloch map is
\begin{equation}
    \Mtar=R(\hat n,\varphi),\qquad \ctar=\bm 0 .
\end{equation}
For example,
\begin{equation}
    \begin{aligned}
    X(\pi):\;\Mtar&=\operatorname{diag}(1,-1,-1),\\
    Z(\pi):\;\Mtar&=\operatorname{diag}(-1,-1,1).
    \end{aligned}
\end{equation}

The learned open-system process is an affine channel on the Bloch ball,
\begin{equation}
    \rout=M\rin+\bm c .
\end{equation}
Because the four probes of the main text are chosen along coordinate
directions, $+\bm e_z$, $-\bm e_z$, $+\bm e_x$, and $+\bm e_y$, the affine
map $(M,\bm c)$ is read directly from their final outputs.  Denoting these
outputs by $\bm s_0,\bm s_1,\bm s_x,\bm s_y$, applying the affine map to
each probe gives
\begin{equation}
\begin{aligned}
    \bm s_0 &= M\bm e_z+\bm c, &
    \bm s_1 &=-M\bm e_z+\bm c,\\
    \bm s_x &= M\bm e_x+\bm c, &
    \bm s_y &= M\bm e_y+\bm c .
\end{aligned}
\label{eq:probe_affine_relations}
\end{equation}
Adding and subtracting the first two outputs gives
\begin{equation}
    \bm c=\frac{\bm s_0+\bm s_1}{2},
    \qquad
    M\bm e_z=\frac{\bm s_0-\bm s_1}{2}.
\end{equation}
The remaining columns are then
\begin{equation}
    M\bm e_x=\bm s_x-\bm c,
    \qquad
    M\bm e_y=\bm s_y-\bm c.
\end{equation}
Thus the channel reconstruction uses only the final propagated probe states.
A nonorthogonal probe set would instead require solving the corresponding
linear system.

\section{Neural-network representation and training procedure}
\label{supp:pinn}

The continuous generator is a three-layer fully connected multilayer perceptron of width 96 with $\tanh$ activations and Xavier/Glorot initialization~\cite{Glorot2010Understanding}; no dropout or batch normalization is used, since the sampled grid is fixed and the Bloch-equation error itself regularizes the representation.  The linear output layer has dimension 14, split into two control heads and four three-component trajectory heads.  Unless otherwise stated, the bounded controls are those of Eq.~\eqref{eq:control_param_main}, without a hard sine envelope.  The trajectory parametrization of Eq.~\eqref{eq:traj_param_main} is used throughout, with the initial-condition envelope
\begin{equation}
    g(t)=1-e^{-t}\qquad(g(0)=0),
\end{equation}
When the sine-envelope pulse prescription is used, the controls are instead multiplied by a hard endpoint envelope,
\begin{equation}
    \begin{aligned}
    \Omega(t)&=\sin\!\left(\tfrac{\pi t}{T}\right)\Omega_{\max}\tanh u_\Omega(t),\\
    \Delta(t)&=\sin\!\left(\tfrac{\pi t}{T}\right)\Delta_{\max}\tanh u_\Delta(t).
    \end{aligned}
\end{equation}

\paragraph{Learnable-duration gradient chain.}
The normalized sampled points $s_i\in[0,1]$ are uniform and fixed; the physical query points $t_i=s_iT$ of Eq.~\eqref{eq:ti_siT_main} are passed directly to the network, so automatic differentiation is taken with respect to the physical time $t$ and the Bloch-equation residual of Eq.~\eqref{eq:dyn_residual_main} needs no manual $1/T$ factor.  The gradient with respect to $\tau$ is nontrivial, because $\tau$ rescales the physical time on which the whole generated process is queried:
\begin{equation}
    \frac{\partial \Ltotal}{\partial \tau}
    =\sum_i\frac{\partial \Ltotal}{\partial t_i}\underbrace{\frac{\partial t_i}{\partial T}}_{s_i}\underbrace{\frac{\partial T}{\partial \tau}}_{T}
    +\left(\frac{\partial \Ltotal}{\partial T}\right)_{\!\rm explicit}\!\frac{\partial T}{\partial \tau}.
    \label{eq:supp_tau_grad_schematic}
\end{equation}
The first term collects all paths through the physical query points (the network outputs, the envelope $g(t_i)$, and the residual); the explicit $T$ dependence covers the final-time readout, the time cost $\LT=T$, and any hard envelope or boundary term.  Thus $\tau$ is coupled to the same forward graph as $\theta$, not optimized in a separate outer loop.

\paragraph{Three pulse-shape prescriptions.}
The prescriptions differ only in which terms of Eq.~\eqref{eq:Lreg_main} are active: \emph{minimal} keeps only the dynamical, gate, and time terms with bounded controls (so the pulse can approach the square-pulse bounded-control reference limit); \emph{regularized} activates the regularization terms; \emph{sine-envelope} replaces the soft boundary penalty by the hard $\sin(\pi t/T)$ envelope above.  Typical rotation-gate weights are $\alpha_{\rm dyn}=1.0$, $\beta_{\rm gate}=10.0$, $\chi_{\rm amp}=10^{-4}$, $\zeta_{\rm smooth}=10^{-4}$, $\eta_{\rm bnd}=10^{-3}$ (hard envelope off), chosen so no single term dominates the initial gradient.

The three regularization terms in Eq.~\eqref{eq:Lreg_main} penalize the pulse amplitude, its rate of change, and its endpoint values:
\begin{equation}
\begin{aligned}
    \mathcal L_{\rm amp}&=\frac{1}{N_s}\sum_i[\Omega(t_i)^2+\Delta(t_i)^2],\\
    \mathcal L_{\rm smooth}&=\frac{1}{N_s}\sum_i[\dot\Omega(t_i)^2+\dot\Delta(t_i)^2],\\
    \mathcal L_{\rm boundary}&=\Omega(0)^2+\Omega(T)^2+\Delta(0)^2+\Delta(T)^2.
\end{aligned}
\end{equation}

\paragraph{Training hyperparameters.}
Rotation gates use Adam~\cite{Kingma2015Adam} at $10^{-3}$ for $\theta$ and a separate $5\times10^{-3}$ for $\tau$ (a single scalar with a different gradient scale), for 3000 steps, with $T_0=0.7$, $\Omega_{\max}=\Delta_{\max}=8.0$, $\lambda_T=1.0$ unless stated otherwise.  Training typically proceeds in three stages, Bloch-equation error first, gate loss next, then a balance of all terms with $T$ stabilizing.  Geometric-gate training uses the same network but a separate path-constraint schedule (\appsecref{supp:geometry}).

\section{Independent validation and evaluation metrics}
\label{supp:validation}

After training we validate the learned controls with an external fourth-order Runge--Kutta integrator, using $N_{\rm RK4}=4000$ uniform steps over $[0,T]$, far more than the $N_s$ points at which the training residual is evaluated, so the validation accuracy is independent of, and much higher than, the training-time residual.  The learned controls leave the training loop only as their values on the $N_s$-point training grid; on the far denser RK4 grid the control fields are obtained by linear interpolation between these stored samples (the network-generated $\Omega(t),\Delta(t)$ are smooth, so the interpolation error is small relative to the RK4 error).  For the discrete-pulse controls the native piecewise-constant bin value is used at each RK4 stage instead.  In both cases the propagation is an independent solve of the Bloch equation under the learned fields, not a re-evaluation of the training residual.

The pointwise trajectory error between the network-internal and the independently propagated trajectories is
\begin{equation}
    \varepsilon_{\rm traj}=\frac{1}{N_pN_s}
    \sum_{k=1}^{N_p}\sum_{i=1}^{N_s}
    \left\|\rr_{\rm PINN}^{(k)}(t_i)-\rr_{\rm val}^{(k)}(t_i)\right\|_2^2.
\end{equation}
The channel is reconstructed from the independently propagated final states, and the process fidelity of Eq.~\eqref{eq:Fproc_main} is evaluated as the entanglement fidelity with respect to the ideal unitary target,
\begin{equation}
    F_{\rm proc}=\frac{1+\operatorname{Tr}(\Mtar^{\mathsf T}M)}{4},
    \label{eq:fproc_ptm_supp}
\end{equation}
equivalently $\operatorname{Tr}(\chi_{\rm tar}\chi_{\rm imp})$ in the unit-trace Choi convention; the affine displacement $\bm c$ enters the gate loss but not this fidelity.  The corresponding average gate fidelity is $F_{\rm avg}=(2F_{\rm proc}+1)/3$~\cite{Nielsen2002AverageGateFidelity}.

\section{Implementation details for geometric-gate tasks}
\label{supp:geometry}

For the geometric-gate extension the learnable-time representation is retained, but the network also outputs reference trajectories and path-constraint terms.  The implemented rotation is reconstructed from three Cartesian probe trajectories $\rr^{(+x)},\rr^{(+y)},\rr^{(+z)}$; for the $Z$ gate $+z$ is fixed, so only $+x$ and $+y$ move.  Two reference trajectories $\rr_\pm(t)$, initialized at $\pm\hat z$, carry the cycle and geometric-phase conditions.  The output dimension is therefore 17: two controls, three probe-trajectory heads, and two reference-trajectory heads.  With $\bm h(t)=(\Omega(t),0,\Delta(t))^{\mathsf T}$, $E_\pm(t)=\bm h(t)\cdot\rr_\pm(t)/2$, and the path losses are
\begin{equation}
\begin{aligned}
    \mathcal L_{\rm geo}&=\tfrac{1}{N_s}\sum_i\!\left[E_+(t_i)^2+E_-(t_i)^2\right],\\[1pt]
    \mathcal L_{\rm cyc}&=\sum_{a=\pm}\|\rr_a(T)-\rr_a(0)\|_2^2,\\[1pt]
    \mathcal L_{\rm pur}&=\tfrac{1}{N_s}\sum_{i,a}\!\left(\|\rr_a(t_i)\|_2^2-1\right)^2,\\[1pt]
    \mathcal L_{\rm orth}&=\tfrac{1}{N_s}\sum_i\!\left[\rr_+(t_i)\cdot\rr_-(t_i)+1\right]^2.
\end{aligned}
\end{equation}
For the representative two-control geometric $Z$ gate we use $\lambda_{\rm dyn}=80$, $\lambda_{\rm gate}=50$, $\lambda_{\rm cyc}=80$, $\lambda_{\rm amp}=\lambda_{\rm smooth}=10^{-4}$, and a geometricity weight held at $\lambda_{\rm geo,init}=10$ for the first 30\% of training and then ramped to $\lambda_{\rm geo}=300$ over the remaining 70\%; the purity and orthogonality terms are auxiliary path-quality penalties on the cycle-constraint scale.  We use a six-layer width-256 network, Adam~\cite{Kingma2015Adam} with cosine annealing from $3\times10^{-4}$ to $10^{-6}$, 20000--30000 steps, $N_s=201$, $\Omega_{\max}=\Delta_{\max}=12.0$, and gradient clipping at norm 5.0.

These choices are driven by the nature of the path constraints.  The geometricity and cycle-closure conditions are pointwise equality constraints, $E_\pm(t)=0$ and $\rr_\pm(T)=\rr_\pm(0)$, and a residual in either shifts the implemented gate as a whole; they must therefore be driven much closer to satisfaction than the soft shape penalties, which is why the geometricity and cycle weights ($\lambda_{\rm geo}=300$, $\lambda_{\rm cyc}=80$) sit orders of magnitude above the amplitude and smoothness weights ($10^{-4}$).  Satisfying the path equalities to the required accuracy while keeping the pulse smooth and bounded is a stiff optimization problem; the larger network and the staged schedule below are the two mechanisms that make it tractable.

The staged schedule matters: if the geometricity term is too strong before the trajectories are dynamically meaningful, the network can reduce the algebraic $E_\pm(t)$ without producing a physically valid controlled path.  Dynamical consistency, final-gate accuracy, and cycle closure therefore dominate early training; the geometricity weight is raised only after the controlled process has organized itself.

The turning-rate diagnostic measures how fast the control direction turns in the $\Omega$--$\Delta$ plane.  With direction angle $\psi(t)$ defined by $(\cos\psi,\sin\psi)=(\Omega,\Delta)/\sqrt{\Omega^2+\Delta^2}$, the exact angular rate is $|\dot\psi|=|\Omega\dot\Delta-\Delta\dot\Omega|/(\Omega^2+\Delta^2)$; we use the regularized form
\begin{equation}
    \kappa(t)=\frac{|\Omega(t)\dot\Delta(t)-\Delta(t)\dot\Omega(t)|}
    {\Omega(t)^2+\Delta(t)^2+\epsilon_\kappa},
\end{equation}
where $\epsilon_\kappa>0$ prevents spurious divergence near vanishing control amplitude, and $\dot\Omega,\dot\Delta,\kappa$ are obtained directly by automatic differentiation.  Large $\kappa(t)$ marks rapid-turning intervals, which the modified geometricity loss weights more heavily,
\begin{equation}
    \mathcal L_{\rm geo}^{\rm turn}=\frac{1}{N_s}\sum_i[1+\beta\kappa_{\rm norm}(t_i)]\left[E_+(t_i)^2+E_-(t_i)^2\right],
\end{equation}
where $\kappa_{\rm norm}$ (normalized by a robust high percentile of $\kappa$ after excluding the gate endpoints, where the envelope drives the amplitude near zero) is a fixed factor, not back-propagated.  In the PINN scan, $\beta\leq30$ improves the geometricity error while keeping $F_{\rm proc}>0.999999$, whereas $\beta=50$ raises the error and gate time; for the piecewise-constant comparison at the matched duration $T=1.5$ we use $\beta=3$, the largest value at which the discrete-pulse baseline remained stable.

The geometric-gate experiments vary two- versus three-control implementations, hard versus soft constraints, fixed- versus learnable-time optimization, and the presence of turning weight.  These are not separate algorithms but different path constraints within the same continuous representation.

\section{Piecewise-constant baseline for geometric-gate comparison}
\label{supp:discrete_pulse}

This section gives the piecewise-constant discrete-pulse baseline used for
the geometric-gate comparison in Fig.~\ref{fig:turn_weighted_loss}.  The
rotation-gate results in the main text no longer use a discrete-pulse
baseline.  For the geometric $Z$ gate we fix the total time to the matched
duration $T=1.5$ and divide $[0,T]$ into $N=200$ bins of width
$\Delta t=T/N$.  The two controls are represented by bin amplitudes
$c_{\Omega,j}$ and $c_{\Delta,j}$, with the same amplitude map and sine
endpoint envelope as the geometric PINN control fields,
\begin{equation}
\begin{aligned}
    \Omega_j&=\Omega_{\max}\sin(\pi t_j/T)c_{\Omega,j},\\
    \Delta_j&=\Delta_{\max}\sin(\pi t_j/T)c_{\Delta,j},
\end{aligned}
\end{equation}
where $t_j$ is the midpoint of bin $j$, $|c_{\Omega,j}|,|c_{\Delta,j}|\leq1$,
and $\Omega_{\max}=\Delta_{\max}=12.0$.  The optimized variables are
therefore the $2N=400$ bin amplitudes.  Trajectories are not optimized:
within each bin the Bloch equation is solved exactly by the matrix
exponential $\Phi_j=\exp(A_j\Delta t)$, where $A_j$ is the Bloch generator
for the constant pair $(\Omega_j,\Delta_j)$.  Composing the $N$ propagators
gives the final probe and reference states.  There is no Bloch-equation
error in the objective, since the trajectories are generated by propagation.

The geometric-gate objective contains the final-gate loss, cycle closure,
geometricity, orthogonality, amplitude, and smoothness penalties,
\begin{equation}
    \begin{aligned}
    \mathcal L={}&\lambda_{\rm gate}\mathcal L_{\rm gate}
    +\lambda_{\rm cyc}\mathcal L_{\rm cyc}
    +\lambda_{\rm geo}\mathcal L_{\rm geo}^{(\beta)}\\
    &{}+\lambda_{\rm orth}\mathcal L_{\rm orth}
    +\lambda_{\rm amp}\mathcal L_{\rm amp}
    +\lambda_{\rm smooth}\mathcal L_{\rm smooth}.
    \end{aligned}
\end{equation}
The propagated $+x,+y,+z$ probes define
$R=[\rr_{+x}(T),\rr_{+y}(T),\rr_{+z}(T)]$, and the propagated reference
paths initialized at $\pm\hat z$ are denoted $\rr_\pm(t_i)$.  The terms are
\begin{equation}
\begin{aligned}
    \mathcal L_{\rm gate}&=\|R-R_Z\|_F^2,\\
    \mathcal L_{\rm cyc}&=\|\rr_+(T)-\hat z\|_2^2
        +\|\rr_-(T)+\hat z\|_2^2,\\
    \mathcal L_{\rm orth}&=\|R^{\mathsf T}R-I\|_F^2
        +[\det(R)-1]^2,\\
    \mathcal L_{\rm amp}&=\frac{1}{N}\sum_j(\Omega_j^2+\Delta_j^2),\\
    \mathcal L_{\rm smooth}&=\frac{1}{N-1}\sum_j
        [(\Omega_{j+1}-\Omega_j)^2+(\Delta_{j+1}-\Delta_j)^2],
\end{aligned}
\end{equation}
where $R_Z=\operatorname{diag}(-1,-1,1)$.  For the turning-weighted
comparison, the geometricity term is
\begin{equation}
    \mathcal L_{\rm geo}^{(\beta)}
    =
    \frac{1}{N+1}\sum_i
    \left[1+\beta\,\kappa_{\rm norm}(t_i)\right]
    \left[E_+(t_i)^2+E_-(t_i)^2\right],
\end{equation}
with $E_\pm(t_i)=\tfrac12[\Omega(t_i)r_{\pm,x}(t_i)+\Delta(t_i)r_{\pm,z}(t_i)]$,
where the piecewise-constant controls are evaluated at the corresponding
node by zero-order hold.  The turning rate is computed from finite
differences of the bin controls by unwrapping the angle of the vector
$(\Omega_j,\Delta_j)$ and normalizing the resulting rate by its maximum.
This is the native finite-bin analogue of the continuous turning diagnostic
in \appsecref{supp:geometry}.  We compare $\beta=0$ with the stable
weighted case $\beta=3$.

The numerical weights are
$\lambda_{\rm gate}=50$, $\lambda_{\rm cyc}=80$,
$\lambda_{\rm orth}=10$, $\lambda_{\rm amp}=10^{-3}$, and
$\lambda_{\rm smooth}=1.0$.  The geometricity weight is staged:
$\lambda_{\rm geo}=10$ in the first L-BFGS-B stage and
$\lambda_{\rm geo}=300$ in the second.  The optimizer is
L-BFGS-B~\cite{Zhu1997LBFGSB} with bounds $[-1,1]$ on every bin
amplitude, function tolerance $10^{-12}$, gradient tolerance $10^{-8}$,
and at most 50 line-search steps.  Each value of $\beta$ uses five
independent restarts: one zero initialization and four random
initializations on $[-0.2,0.2]$.  The first stage allows up to 250
iterations and the second up to 350 iterations; the best restart by
independent average gate fidelity is kept.  The final controls are then
validated by the same independent RK4 procedure as the PINN controls, with
4000 steps and zero-order hold on the piecewise-constant controls.  This
tests a finite piecewise-constant implementation in which difficult path
intervals are weighted more heavily; it is not a statement about all
conventional optimal-control parametrizations.

\bibliography{refs}

\end{document}